# Thermal and Size Effects in Ferroelastic Domains by Machine Learning

*Luka Geddis Zellmann\*; Sumner B. Harris, John R. R. Scott, Yi-Chieh Yang, Joerg R. Jinschek, Rama K. Vasudevan, Miryam Arredondo\**

L. Geddis Zellmann, J. R. R. Scott, M. Arredondo

School of Mathematics and Physics, Queen's University Belfast, Belfast, UK

Email: lgeddiszellmann01@qub.ac.uk, m.arredondo@qub.ac.uk

S. B. Harris, R. K. Vasudevan

Center for Nanophase Materials Sciences (CNMS), Oak Ridge National Laboratory (ORNL), Oak Ridge, USA

Y.-C. Yang,  J. R. Jinschek

National Center for Nano Fabrication and Characterization (DTU Nanolab), Technical University of Denmark, Kgs. Lyngby, Denmark

Funding: EPSRC PIADS studentship (grant no: EP/S023321/1), ORNL CNMS user project (proposal number: CNMS2024-B-02660)

Ferroelastic domain walls (DWs) underpin key functionalities in complex oxides. In free-standing ferroic thin films, where elastic interactions are highly thickness dependent, understanding DW behavior across length scales and external stimuli is crucial. A thickness-dependent monopolar-dipolar crossover in elastic DW behavior has been reported, however, how temperature influences this regime remains unexplored. Here, $LaAlO_3$ thin films spanning the dipolar (< 200 nm) and crossover (200–300 nm) regimes are investigated using in situ heating scanning transmission electron microscopy (STEM) and a machine learning-driven image analysis approach. By tracking DW curvature and density from above $T_C$ (~550 °C) to room temperature (RT), a distinct interplay between temperature and thickness is identified. In the dipolar regime, DWs are mobile and curved near $T_C$ and gradually freeze upon cooling, consistent with the well-known temperature freezing regime. In contrast, within the crossover regime DWs are nearly static, with minimal reconfiguration through cooling and curvature an order of magnitude lower at RT. These results map the evolution of DWs across the thermally driven super-elastic to freezing regimes, revealing how thickness and temperature govern DW morphology and dynamics, and providing insight relevant for domain engineering in free-standing oxide thin films.

## 1. Introduction

Ferroelastic materials, like other ferroics, exhibit regions of uniform order parameter orientation, known as domains and separated by domain walls (DWs), which can be switched by the application of external fields. Ferroelastic domains originate from spontaneous strain, with DWs accommodating lattice distortions between differently oriented domains. Beyond serving as passive strain boundaries, DWs are recognized as active functional entities with emergent physical properties[1] such as polarity[2,3] and conductivity.[4,5] This concept is referred to as "domain boundary engineering"[6] where the structure, properties and dynamics of DWs can be exploited as nanoscale templates for devices.[7,8] This approach offers routes to emergent properties with potential for novel applications in neuromorphic computing,[9,10] multiferroic devices,[11,12] and adaptive electronics.[13]

In ferroelastics such as the model system $LaAlO_3$ (LAO), DW dynamics are governed by strain mediated interactions, defect pinning, and long-range elastic forces.[14] Salje's theoretical models established that kinks (atomic steps contained within DWs) interact via long range strain forces, behaving as elastic monopoles in bulk or dipoles in thin film-like systems, with distance dependencies of $d^{-1}$ and $d^{-2}$ respectively, and predicted a size-dependent transition in the form of a monopolar–dipolar crossover regime around a critical thickness.[15] Recent studies using scanning transmission electron microscopy (STEM) have confirmed these predictions experimentally, revealing that thickness significantly influences domain configuration and identifying the crossover regime.[16]

While these studies revealed the static, thickness-dependent domain structure, the thermally driven dynamics within the crossover regime are still largely unexplored. In bulk ferroelastics, temperature plays a critical role, controlling wall mobility and defect pinning through two distinct temperature regimes: a super-elastic regime at elevated temperatures, where DWs are highly mobile and responsive, and a freezing regime where DWs become pinned by defects and mobility is suppressed.[17] Temperature dependent studies in bulk LAO demonstrated that geometrical features, such as aspect ratio, strongly influence domain and DW evolution.[18]

However, how thermal activation and size effects couple, and whether temperature can be leveraged to tune DW morphology and mobility remains unknown. This coupling would be particularly significant at the nanoscale, where effects such as DW curvature and junction formation are amplified.

Here, we investigate how temperature influences DW response across thickness in free-standing LAO films, spanning the monopolar-dipolar regimes. Using in situ heating STEM combined with machine learning–based image segmentation, we quantify changes in DW density and curvature cooling from above the phase transition ($T_C$). This study demonstrates a strong coupling between thickness and temperature, revealing that these can control DW curvature, density, and dynamic response, establishing a foundation for domain boundary engineering in free-standing oxide membranes and new routes toward adaptive domain-engineered devices.

## 2. Results and Discussion

Free-standing LaAlO$_3$ (LAO) samples with thicknesses between 160 nm and 300 nm were fabricated by focused ion beam (FIB) milling and lifted out from bulk single crystals, following the procedure described in ref. [16]. The [100]$_{pc}$ oriented samples were investigated by in situ STEM during controlled heating and cooling from room temperature (RT) through $T_C$ (~550 °C). Using a DENSsolutions Climate in situ holder, all samples were heat-cycled from RT to 600 °C and back at 0.33 °C s$^{-1}$ (see Methods). STEM dark field (STEM-DF) images were acquired at fixed temperature intervals during cooling. **Figure 1** displays representative STEM-DF images at selected temperatures, with the full dataset provided in the Supporting Information (Figure S1, S2 & S3).

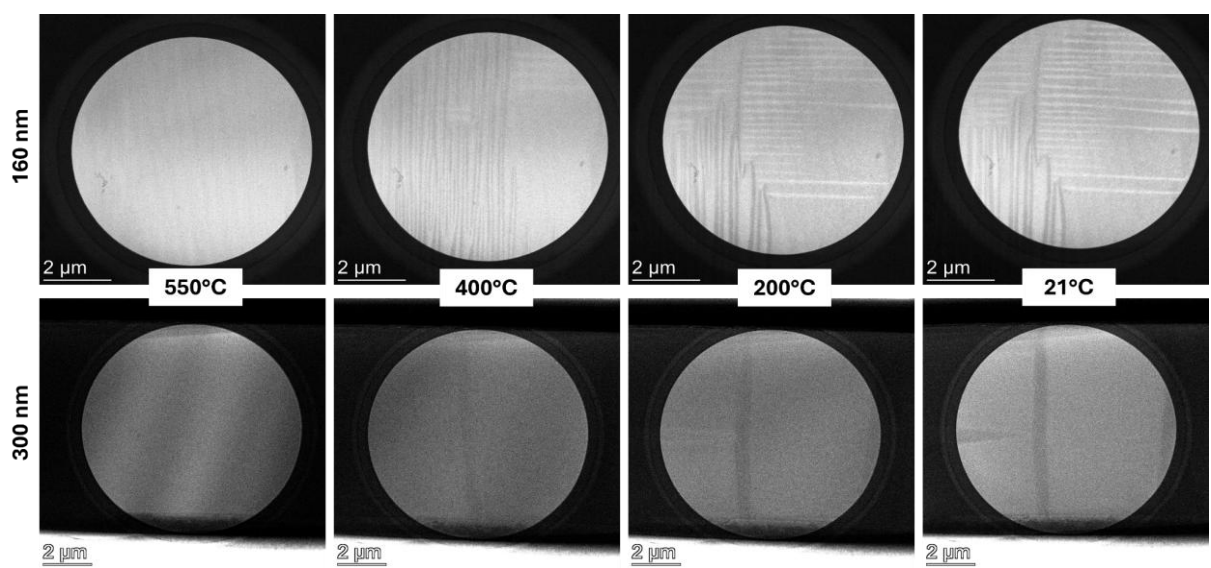

**Figure 1.** Evolution of the domain wall (DW) structure as a function of temperature, cooling from $T_C$. Representative scanning transmission electron microscopy dark field (STEM-DF) images showing domain structure in 160 nm and 300 nm thick LaAlO$_3$ samples during cooling from 550 °C to room temperature. The 160 nm sample displays a dense, curved domain configuration while the 300 nm sample shows a sparse and largely immobile configuration throughout.

For this work, 10 samples were imaged at 50 °C temperature intervals between RT and $T_C$. Manual analysis of large in situ datasets, particularly STEM images, is inherently challenging due to the high volume of images and is a very time-consuming task that can be plagued with human error. Machine learning-based image segmentation offers a promising alternative method, enabling a more accurate analysis in a vastly reduced timeframe. Fully convolutional neural networks (FCNNs) such as U-Net[19] have become popular within fields such as biomedical science,[20] and have recently shown increasing use in materials science.[21,22] These models are capable of semantic segmentation (the classification of each individual pixel in an image) while requiring only a small set of hand-labelled training data.

Here, a machine-learning model was trained to identify DWs, enabling a more robust quantification of DW density and curvature as a function of temperature. The model is based on U-Net[19] with a ResNet-34 encoder;[23] the workflow is illustrated in **Figure 2**. A representative subset of images (three images per sample, ≈25% of the dataset) was randomly selected and manually annotated pixel-wise into three classes: DWs, domains, and background (regions outside the transparent window of the MEMS in situ chip), using the open source data annotation software Label Studio.[24] Of these annotated images, 75% were used for training and the remaining 25% were reserved as a ground-truth validation set.

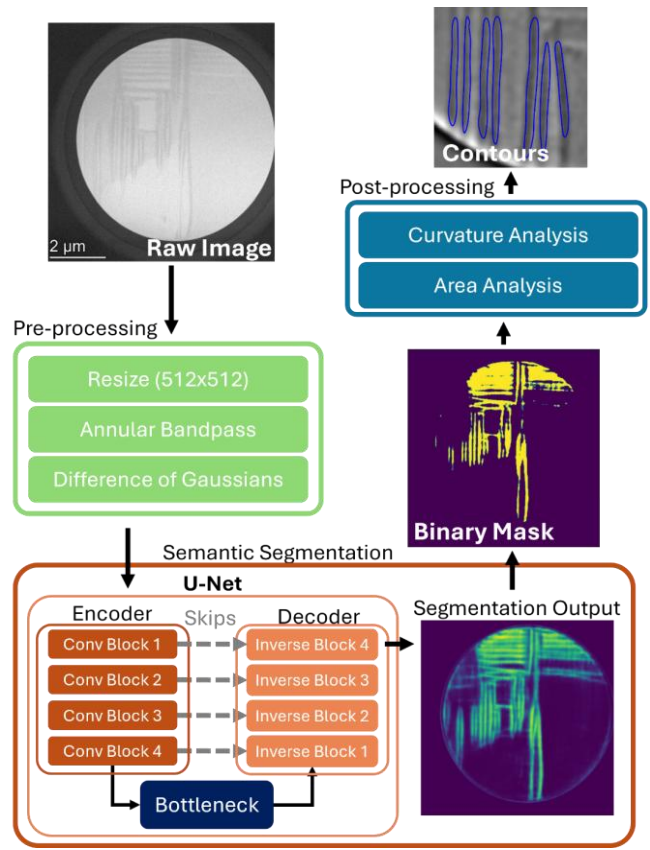

**Figure 2.** Machine learning workflow used to calculate domain wall (DW) curvature and area measurements from scanning transmission electron microscopy dark field (STEM-DF) images. Raw STEM-DF images undergo filtered preprocessing to eliminate image artifacts before semantic segmentation is carried out using U-NET. A binary mask is produced from the U-NET output and used to calculate domain wall area and curvature.

The model performs semantic segmentation to classify each pixel as DW, domain or background (Figure S4 shows representative domain wall predictions output by the trained model). Predicted DW masks were used to compute DW area (a representation of DW density), and contiguous DW pixels in representative cropped regions of the image were fitted with splines to extract local curvature. DW area and curvature were averaged across all samples with a given thickness at each temperature interval to yield mean values for each sample thickness set.

The machine learning analysis revealed pronounced thickness-dependent differences in DW morphology and behavior during cooling. **Figure 3** compares the mean DW area fraction and mean log(curvature) during cooling for all sample thicknesses studied. For thin samples (160–180 nm), the DW area fraction increases rapidly just below $T_C$, reaching a maximum between 300–400 °C before gradually decreasing and plateauing at ≈200 °C. In contrast, samples in the 200–300 nm crossover regime exhibited a nearly constant, low DW area across all temperatures, with small, discrete increases in DW area that correspond to the formation of individual DWs, indicating strongly suppressed nucleation and mobility.

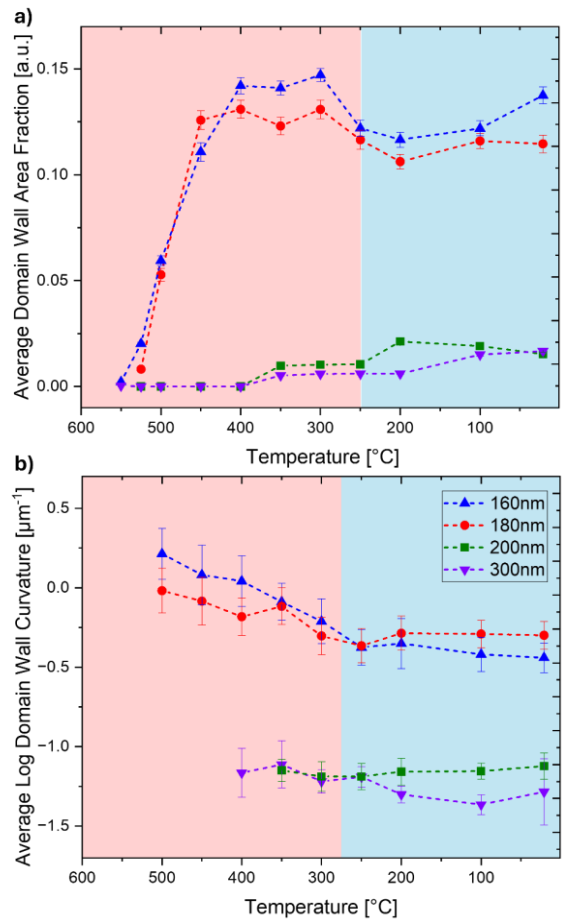

**Figure 3.** Temperature-dependent evolution of average a) domain wall (DW) area fraction and b) log of DW curvature during cooling from $T_C$ across thicknesses. The red shading represents the super-elastic temperature regime while the blue shading represents the freezing regime. Uncertainties derived using bootstrapping, uncertainties for 200 nm and 300 nm domain wall area fraction ~1×10$^{-4}$.

A similar trend was observed for DW curvature. In the thinner samples, DWs exhibit a high curvature value of ~1-1.5 µm$^{-1}$ near $T_C$, which decreases continuously on cooling and plateaus at ≈250 °C with curvatures in the range ~0.3-0.5 µm$^{-1}$ (Figure S6 shows a comparison of DW area fraction and curvature for 160-180 nm samples). Thicker samples displayed lower curvature by approximately an order of magnitude throughout the entire cooling cycle, again indicating suppressed nucleation, pinned walls and mobility.

These results extend previous room temperature studies [16] that established a monopolar-dipolar elastic crossover near 200 nm in thickness. Importantly, these trends demonstrate that DW mobility, curvature, and density are jointly governed by the interplay between sample thickness and temperature. The thinner samples maintain high curvature and dynamic DW motion well below $T_C$, whereas thicker samples, falling into the crossover regime, exhibit minimal DW curvature, reflecting a transition to a monopolar dominated elastic regime and effectively freeze early, even at temperatures characteristic of the super-elastic regime.

This suppression of DW mobility aligns with the theoretical prediction[15] that elastic interactions transition from dipolar ($\propto d^{-2}$) to monopolar ($\propto d^{-1}$) character with increasing thickness. In the thin dipolar limit, DWs interact weakly and can bend to minimize local strain, giving rise to highly curved and mobile DWs. As thickness increases, monopolar behavior amplifies long-range elastic restoring forces, pinning DWs, penalizing curvature and favoring straight DWs. These experimental findings align with theoretical observations and, to the best of our knowledge, provide the first evidence of the temperature-driven response for this thickness regime.

The temperature-dependence of DW mobility observed here closely follows the super-elastic and freezing regimes reported for bulk LAO,[17,25] where dynamic mechanical analysis showed that LAO exhibits strong anelastic softening between $T_C$ (~545 °C) and 300 °C, corresponding to the super-elastic regime, followed by a dramatic mobility decrease below ~250 °C (freezing regime), where defect pinning dominates. The evolution of the DW curvature and area fraction observed in the thinnest samples reproduces this trend on the nanoscale. As seen in Figure 3, DWs are highly mobile and curved between 550 °C and 300 °C, consistent with super-elastic behavior, and become increasingly pinned below 250 °C, marking the onset of the freezing regime. In contrast, thicker crossover samples remain largely static through cooling, implying that size-dependent elastic suppression can mimic freezing-like behavior even above 300 °C (nominally super-elastic temperatures).

Interestingly, in the crossover regime (200–300 nm), domains appeared only at temperatures of ≈400 °C and below, suggesting a subtle thickness-induced shift in $T_C$. Such shifts may arise from strain confinement or finite-size effects that alter the balance of elastic and entropic contributions to the free energy. To evaluate this apparent shift in transition temperature and the possible effect of thermal gradients or mechanical bulging from the in-situ heating geometry, COMSOL finite-element simulations were performed (**Figure 4**). Supporting Information, Figure S7 and Table S1 show model geometry and sample parameters used.

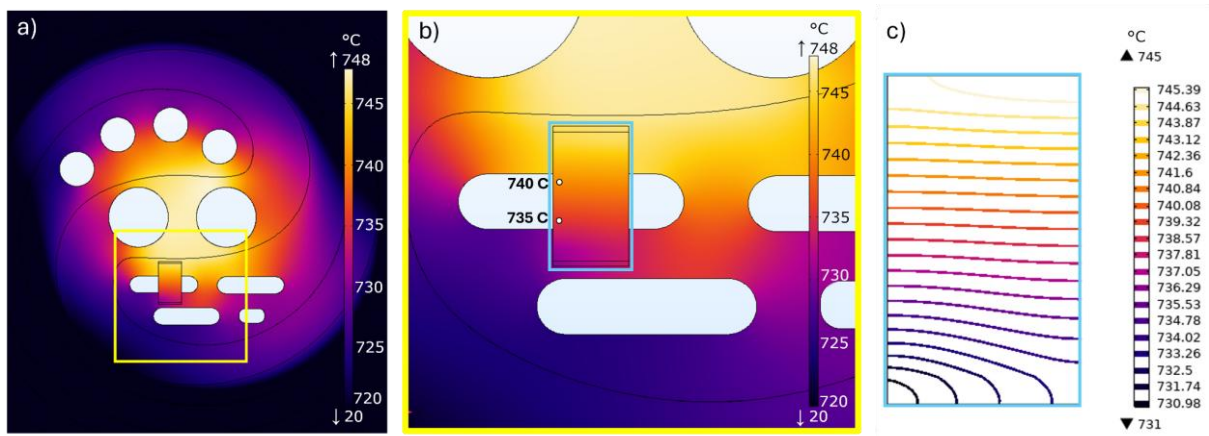

**Figure 4.** – COMSOL simulation of thermal gradients in the in situ MEMS heating geometry. a) Temperature distribution across the microheater window at 747 °C (1.2 V input), indicating a difference of 10±3 °C between the inner and outer windows. b) Temperature variation across an LaAlO$_3$ sample positioned on a window, showing a thermal gradient of 5±3 °C from one edge of the window to the other. c) Thermal gradient across the sample's width (including areas away from the window); the total gradient is ~12 °C.

These simulations show a temperature gradient across the MEMS window of ~5 ± 3 °C and ~12 ± 3 °C across the full sample (Figure 4), and ~5 μm out-of-plane bulging (see SI, Figure S8) at set temperatures as high as 747 °C. Although the experimental heating cycles were conducted at set temperatures up to 600 °C, the COMSOL simulations in Figure 4 were run at 747 °C to place the sample under a worst-case thermal load. If temperature gradients and bulging remain small and uniform even at this higher simulated temperature, then thermal non-uniformity would be equal or lower under the actual experimental conditions. Figure S8 further evaluates membrane deformation under heating, matched directly to the experimental MEMS chip layout, showing that even under elevated temperatures (up to 747 °C) the total out-of-plane bulging has minor variation between different thickness. However, the shift in $T_C$ was only observed in thicker samples (200-300 nm) within the crossover regime. This strongly suggests that this change in $T_C$ is intrinsic and thickness-driven, not an artefact.

The curvature and density variations observed here have direct implications for nanoscale functionality. Vasudevan et al. demonstrated that curvature modulates conductivity in $BiFeO_3$, with local DW bending enhancing carrier accumulation.[26] Similarly, studies in $ErMnO_3$ revealed that DW bending introduces highly charged DW segments that modulate local electric fields, resulting in conductivity variations at the nanoscale.[27] These studies establish a clear link between DW geometry and charge transport. By analogy, curvature-dependent strain fields in ferroelastics such as LAO could influence ionic or phonon transport, suggesting a route to further tune local strain for thermal or electromechanical properties. This highlights curvature, controlled through both thickness and temperature, as a powerful parameter for designing emergent functionality in ferroelastic membranes.

### 3. Conclusion

This study demonstrates that temperature and thickness act as coupled parameters that can govern DW morphology and mobility in free-standing ferroelastics.

Overall, thicker samples (200–300 nm) exhibited simpler domain configurations, with reduced domain wall (DW) density and fewer junctions, with straighter DWs. Thinner samples (<200 nm) displayed more complex configurations, with a higher density of DWs, higher number of junctions, and unique to this thickness range, DW curvature that is increased at temperatures near $T_C$. Thinner, dipolar samples remain mobile throughout cooling, while thicker, crossover regime samples experience early freezing.

This work connects size-dependent elastic interactions with temperature-driven anelasticity, offering a basis for designing ferroelastic membranes with tailored dynamic responses. Future studies combining time-resolved in situ diffraction, acoustic spectroscopy, and phase-field simulations could further elucidate how elastic softening and defect pinning collectively define DW dynamics and functional properties at the nanoscale.

### 4. Experimental Section/Methods

Free-standing samples of ≈7 μm × 12 μm, with thicknesses ranging from 160 to 300 nm were prepared using a TESCAN Lyra 3 dual beam FIB/SEM by standard FIB milling techniques. These samples were fabricated to have posts on either side, similar to previous in situ heating TEM studies[28] to prevent direct contact with the electron transparent window during heating and resemble free-standing conditions, and were placed on an in situ heating MEMS chips via conventional ex situ lift out. Fully detailed in Ref. [16]. 10 samples were analysed in this study : 3 samples of 160 nm, 5 samples of 180 nm, 1 sample of 200 nm, and 1 sample of 300 nm.

Scanning transmission electron microscopy (STEM) images were recorded using a dark field STEM (STEM-DF) detector on an FEI TALOS F200 G2 at 200 kV, with a dwell time of 20 µs.

All samples were heat cycled from room temperature to 600 °C (above $T_C$ ~550 °C) and back at a rate of 0.33 °C s$^{-1}$, using a DENSsolutions Climate system with an open cell configuration, as described elsewhere.[16]

Image analysis was carried out using a U-NET[19] model with a ResNet-34[23] encoder. The model was trained using manually labelled images, 33 of which were used as a training set and another 9 images used as the ground-truth validation set. All manual labelling was performed using the open-source data annotation software Label Studio.[24] Prior to training, images were pre-processed with spatial filters (difference-of-Gaussians and annular bandpass) to suppress imaging artefacts as much as possible, most notably bending contours. Bending contours are large regions of alternating dark and light contrast that arise from thermal expansion of the samples and the in situ MEMS chip; the applied filters minimized these features to reduce their influence on model performance. The loss function was the mean of binary cross-entropy (BCE) and Dice loss. Model hyperparameters were: learning rate $\eta = 1 \times 10^{-4}$, batch size $|B| = 2$, and class-weighting for DW:domain:background = 60:30:10. The model achieved a final test loss of 0.057 (see training curve in Figure S5).

## Acknowledgements

L.G.Z. and M.A. acknowledge the UK EPSRC and SFI Centre for Doctoral Training in Photonic Integration and Advanced Data Storage (PIADS) program for the sponsorship of PhD studentship (grant no: EP/S023321/1). Machine learning model development was conducted as part of a user project at the Center for Nanophase Materials Sciences (CNMS), which is a US Department of Energy, Office of Science User Facility at Oak Ridge National Laboratory (proposal number: CNMS2024-B-02660).

## Conflict of Interest

The authors disclose no conflict of interest.

## Data Availability

The code used for this study can be accessed here: https://github.com/lged01/LaAlO3-Segmentation-Network. The training data used, as well as the final model, is available here: https://www.dropbox.com/scl/fo/a86c23bf74xkc9l6lqdd6/AMJ66EC8MvMy0o8Rc6VDDu4?rlkey=26tx9n36njy09sgjo2osb6o0q&st=9i9eymcl&dl=0.

**Supporting Information for "Thermal and Size Effects in Ferroelastic Domains by Machine Learning"**

*Luka Geddis Zellmann\*, Sumner B. Harris, John R. R. Scott, Yi-Chieh Yang, Joerg R. Jinschek, Rama K. Vasudevan, Miryam Arredondo\**

L. Geddis Zellmann, J. R. R. Scott, M. Arredondo

School of Mathematics and Physics, Queen's University Belfast, Belfast, UK

Email: lgeddiszellmann01@qub.ac.uk, m.arredondo@qub.ac.uk

S. B. Harris, R. K. Vasudevan

Center for Nanophase Materials Sciences (CNMS), Oak Ridge National Laboratory (ORNL), Oak Ridge, USA

Y.-C. Yang, J. R. Jinschek

National Center for Nano Fabrication and Characterization (DTU Nanolab), Technical University of Denmark, Kgs. Lyngby, Denmark

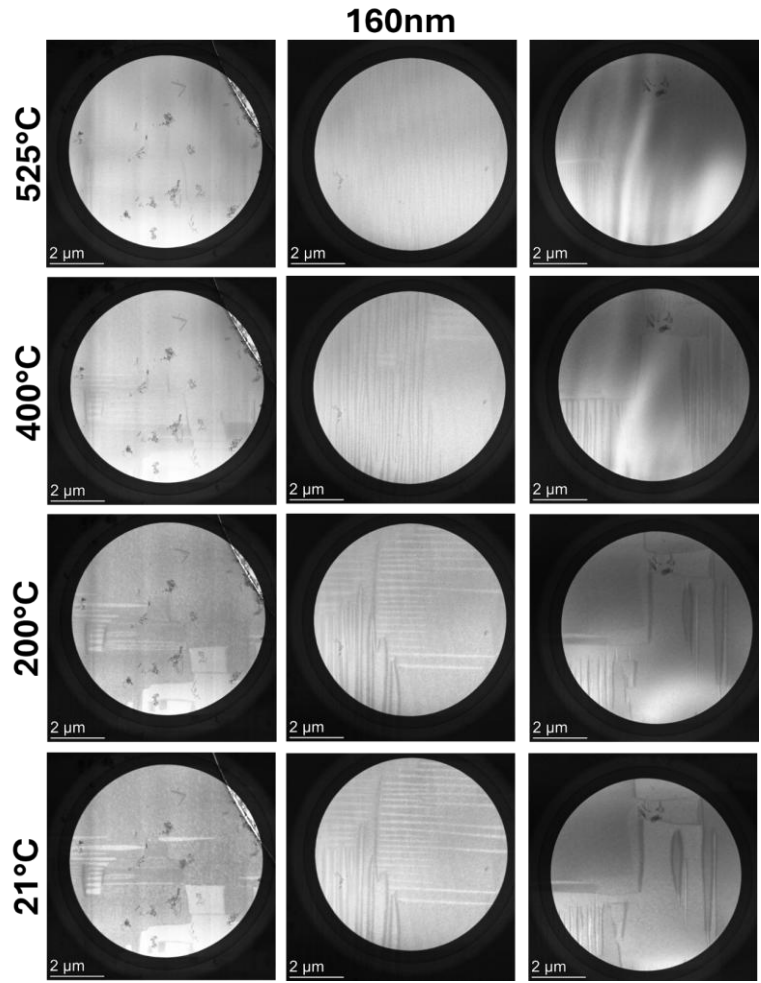

**Figure S1**. Scanning transmission electron microscopy dark field (STEM-DF) images of three 160 nm free-standing LaAlO$_3$ samples during cooling from 550 °C to room temperature, showing dense, curved, and highly mobile domain walls near T$_C$ that gradually straighten and freeze as temperature decreases.

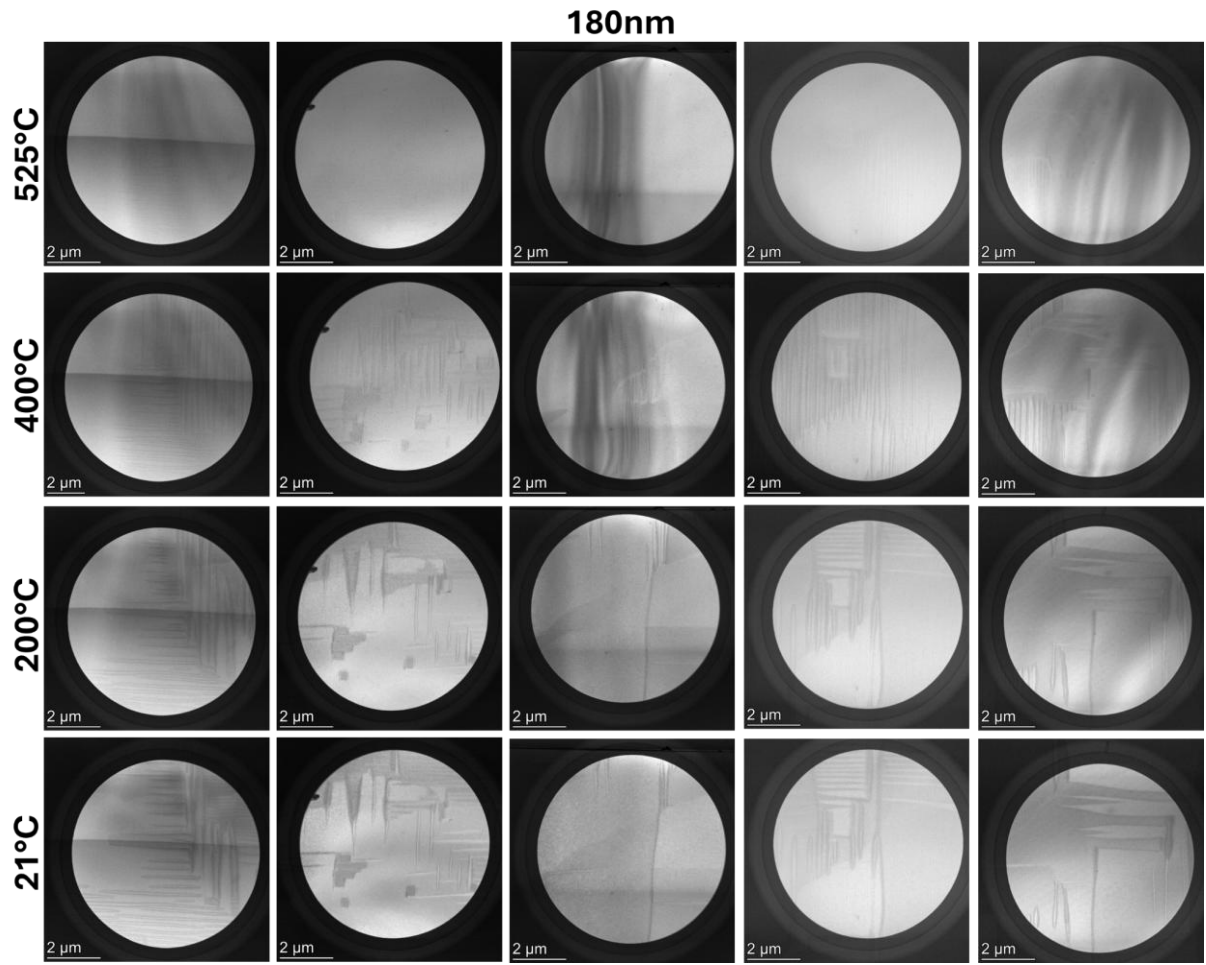

**Figure S2**. Scanning transmission electron microscopy dark field (STEM-DF)images of five 180 nm free-standing LaAlO$_3$ samples during cooling from 550 °C to room temperature, revealing the progressive straightening and reduced density of domain walls as the system transitions from the super-elastic to freezing regime in the crossover thickness range.

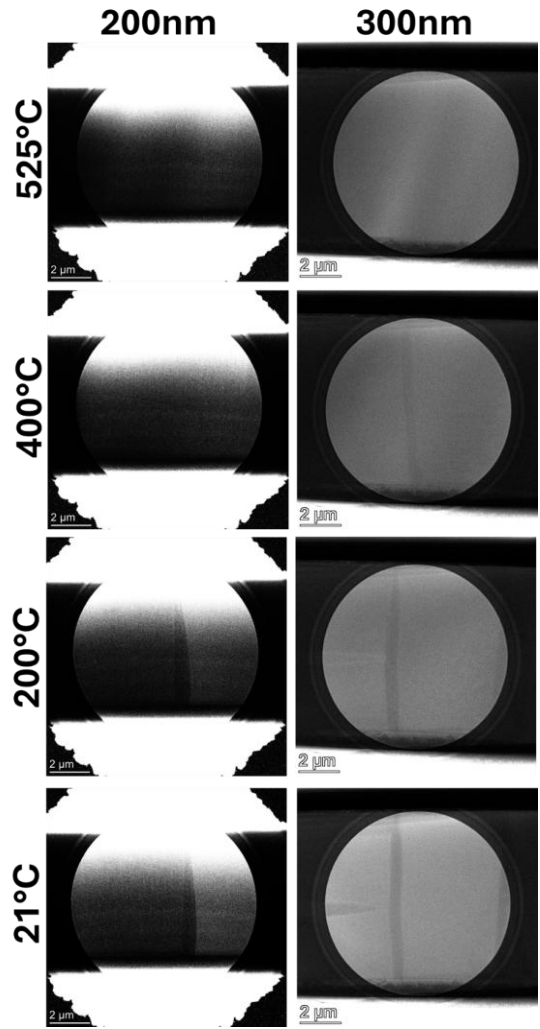

**Figure S3**. Scanning transmission electron microscopy dark field (STEM-DF) images of 200 nm and 300 nm free-standing LaAlO$_3$ samples during cooling from 550 °C to room temperature, showing static, low-curvature domain walls.

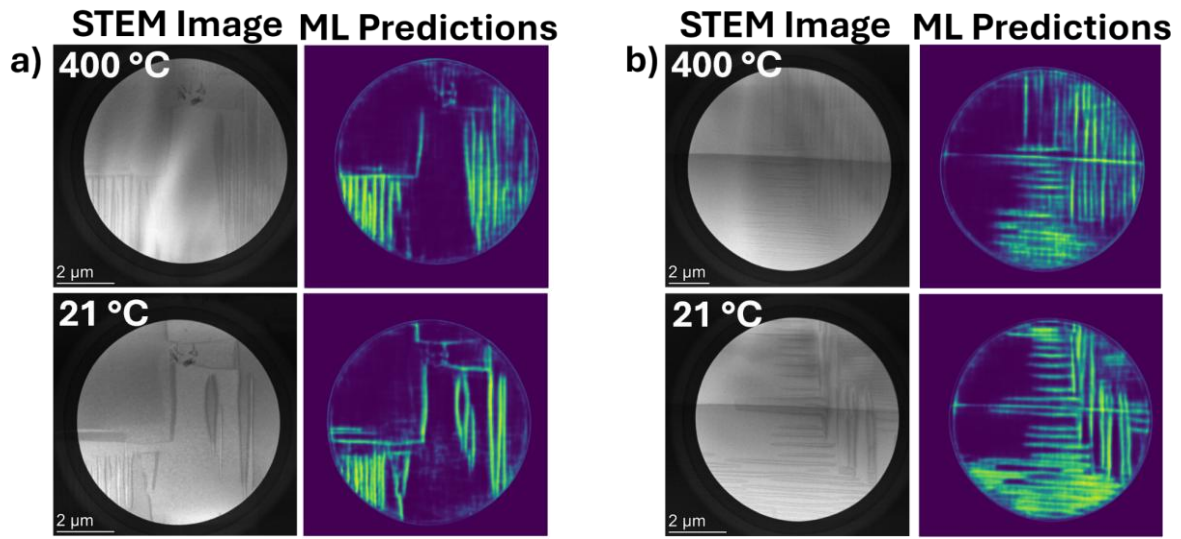

**Figure S4.** Scanning transmission electron microscopy (STEM) images of an a) 160 nm and b) 180 nm $LaAlO_3$ sample at 400 °C and 21 °C, alongside segmentation outputs from the trained U-NET model showing the predicted positions of domain walls (DWs). Even when contrast is significantly affected by bending contours, as in the 400 °C images, the model is able to identify DWs accurately.

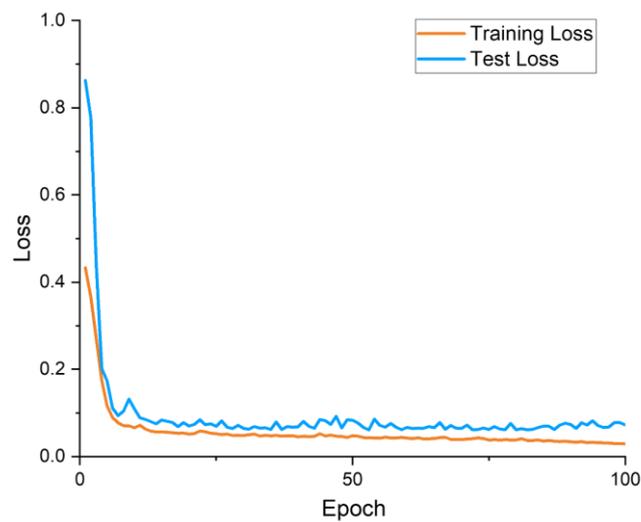

**Figure S5.** Training curve showing the evolution of the combined binary cross-entropy (BCE) and Dice loss during model optimization for domain wall segmentation. The steady decrease and plateau of the loss function indicate stable convergence and robust performance of the U-Net (ResNet-34 encoder) model.

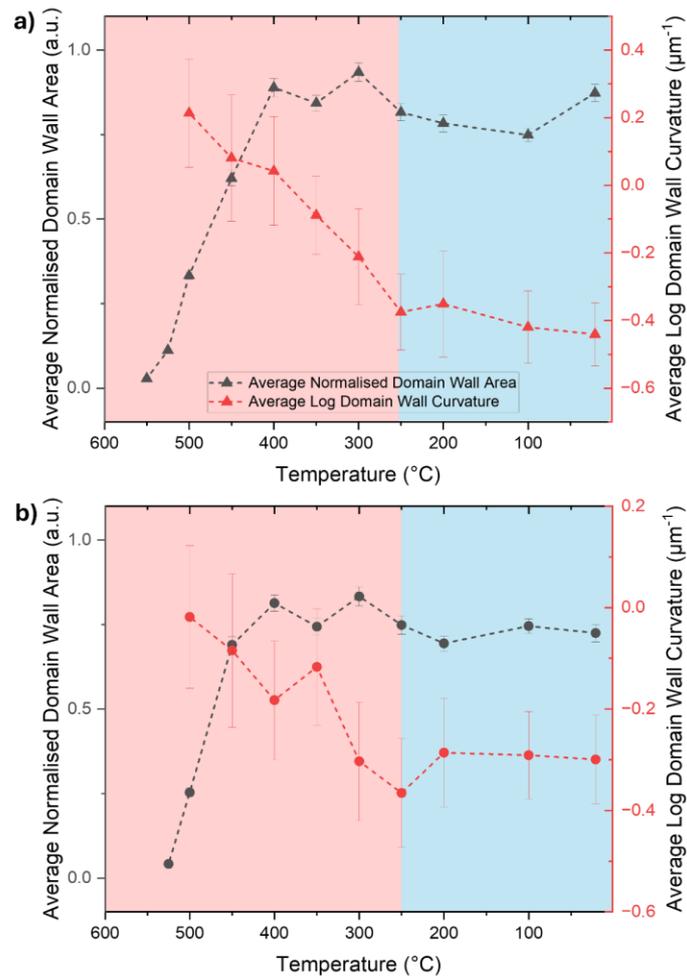

**Figure S6.** Temperature-dependent evolution of normalized domain wall area and curvature in (a) 160 nm and (b) 180 nm free-standing LaAlO$_3$ (LAO). Both samples exhibit increased domain wall density and high curvature near T$_C$, consistent with super-elastic mobility, followed by progressive straightening and reduced area as the system cools into the freezing regime below ~250 °C. The shaded regions indicate the super-elastic (red) and freezing (blue) regimes.

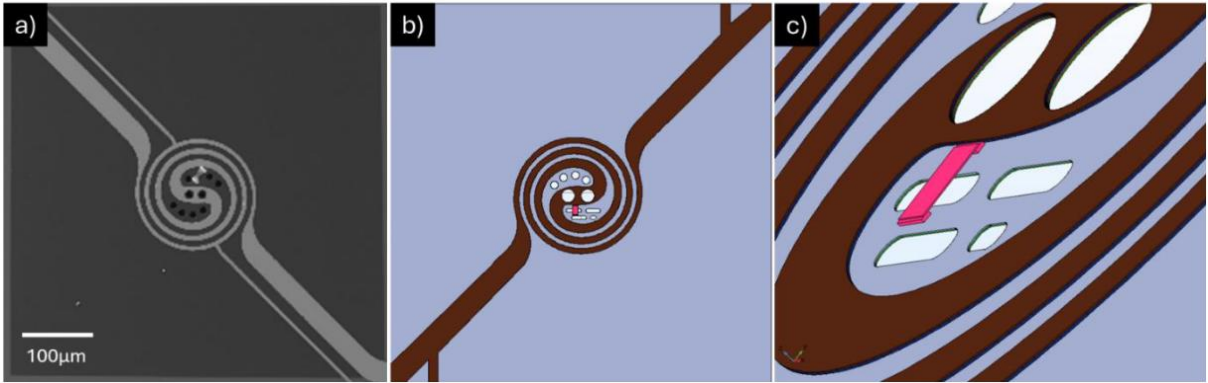

**Figure S7.** COMSOL model based on scanning electron microscopy (SEM) measurements of a free-standing LaAlO$_3$ (LAO) sample on a DENSsolutions in situ heating chip. (a) SEM image of the experimental configuration, (b) corresponding COMSOL geometry used for simulations (based on a DENSsolutions Wildfire chip), and (c) modelled LAO sample with identical dimensions to the experimental samples.

| Property | Value | Unit |
|---|---|---|
| Coefficient of Thermal Expansion | 10x10$^{-6}$ | 1/K |
| Heat Capacity at Constant Pressure | 730 | J(kg*K) |
| Density | 6.56 | g/cm$^3$ |
| Thermal Conductivity | 10 | W(m*K) |
| Young's Modulus | 400x10$^9$ | Pa |
| Poisson's Ratio | 0.22 | No Units |
| Electrical Conductivity | 0 | S/m |
| Relative Permittivity | 5.7 | No Units |

**Table S1.** LaAlO$_3$ parameters used in the COMSOL simulations.

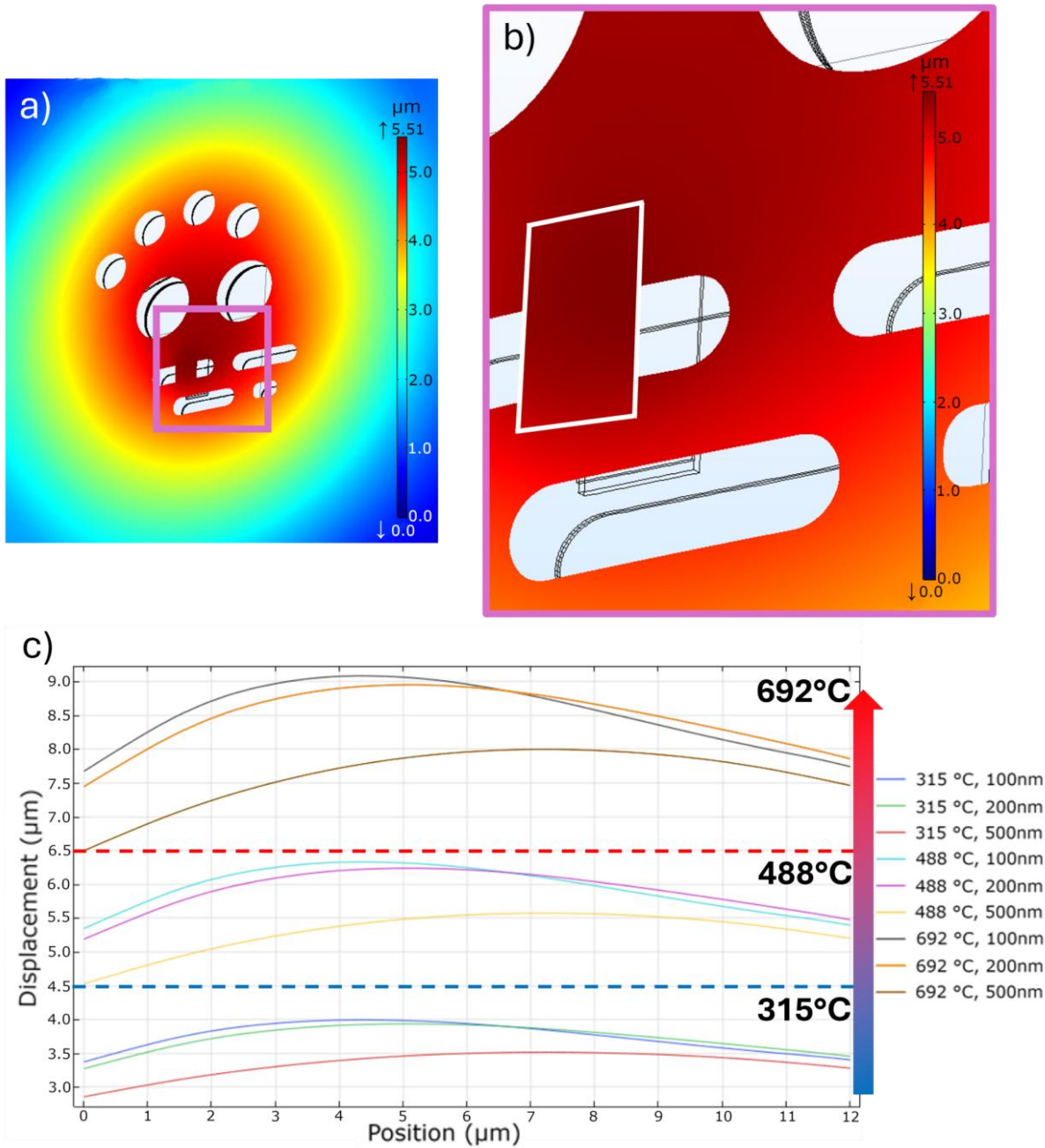

**Figure S8.** COMSOL simulation of the bulging experienced by the a) microheater membrane and b) LAO sample (outlined in white) at 747°C (input = 1.2V). The overall bulging experienced by the sample is approximately 5μm. c) Out-of-plane displacement due to bulging across the lengths of the sample for samples of varying thickness at several temperatures. Bulging is greater at higher temperatures and thinner samples experience slightly greater bulging than thicker samples at the same temperature.